\documentclass[final,3p,times]{elsarticle}

\usepackage{graphicx}
\usepackage{amssymb}
\journal{Nuclear Physics A}

\begin{document}

%%%%%%%%%%%%%%%%%%%%%%%%%%%%%%%%%%%%%%%%%%%%%%%%%%%%%%%%%%%%%%%%%%
\begin{frontmatter}

\title{In-medium effects on hypernuclear formation}

\author[a]{T.~Gaitanos}
\author[b,c]{A.B.~Larionov}
\author[a]{H.~Lenske}
\author[a]{U.~Mosel}
\author[a]{A.~Obermann}

\address[a]{Institut f\"ur Theoretische Physik, Universit\"at Giessen, Germany}
\address[b]{Frankfurt Inst. for Adv. Studies, J.W. Goethe-Universit\"at, Frankfurt, Germany}
\address[c]{Russian Research Center, Kurchatov Institute, Moscow, Russia}
%\address{email: Theodoros.Gaitanos@theo.physik.uni-giessen.de}
%%%%%%%%%%%%%%%%%%%%%%%%%%%%%%%%%%%%%%%%%%%%%%%%%%%%%%%%%%%%%%%%%%%
\begin{abstract}
We study strangeness dynamics in reactions relevant for the formation of 
multi-strange hypernuclei. In particular, we discuss the in-medium effects 
on elementary hyperon-nucleon ($YN$) channels, which are relevant for the 
production of hypernuclei at PANDA. The results indicate strong in-medium 
effects on $YN$-scattering, which might be important for hypernuclear 
studies at PANDA.
\end{abstract}
%******************************************************************


\end{frontmatter}

%\end{quote}}
%\date{\today}
%\maketitle
%\setpapersize{A4}
      %    Width of rule between columns.
%%%%%%%%%%%%%%%%%%%%%%%%%%%%%%%%%%%%%%%%%%%%%%%%%%%%%%%%%%%%%%%%%%%
%%%%%%%%%%%%%%%%%%%%%%%%%%%%%%%%%%%%%%%%%%%%%%%%%%%%%%%%%%%%%%%%%%%
%%%%%%%%%%%%%%%%%%%%%%%%%%%%%%%%%%%%%%%%%%%%%%%%%%%%%%%%%%%%%%%%%%%
%           BEGIN OF TEXT                 %
%%%%%%%%%%%%%%%%%%%%%%%%%%%%%%%%%%%%%%%%%%%%%%%%%%%%%%%%%%%%%%%%%%%
%%%%%%%%%%%%%%%%%%%%%%%%%%%%%%%%%%%%%%%%%%%%%%%%%%%%%%%%%%%%%%%%%%%
%%%%%%%%%%%%%%%%%%%%%%%%%%%%%%%%%%%%%%%%%%%%%%%%%%%%%%%%%%%%%%%%%%%

%%%%%%%%%%%%%%%%%%%%%%%%%%%%%%%%%%%%%%%%%%%%%%%%%%%%%%%%%%%%%%%%%%%
\section{Introduction}
\label{sec1}
%%%%%%%%%%%%%%%%%%%%%%%%%%%%%%%%%%%%%%%%%%%%%%%%%%%%%%%%%%%%%%%%%%%

The study of hypernuclei is intimately related to many fundamental 
aspects of nuclear and hadron physics~\cite{ref1}. Spectroscopy can 
be at best performed with hyperons, since single hyperons are not Pauli 
blocked in nucleonic matter and can provide particular clean 
information on single-particle spectroscopy. 
The investigation of hypernuclei can provide information also on 
the still less known hyperon-nucleon and hyperon-hyperon 
interactions. For the latter the production of double-$\Lambda$ 
hypernuclei is important, which is one of the main projects of the 
PANDA Collaboration~\cite{ref1a}.

Recently, we have studied the formation of $S=-1$ and $S=-2$ hypernuclei 
in various reactions~\cite{ref2} within a covariant transport 
model~\cite{ref3}. Here we focus on the discussion of the production 
mechanisms of double-$\Lambda$ hypernuclei at PANDA. In particular, 
we examine possible in-medium dependences 
of hyperon-nucleon interactions important for the formation of hypernuclei 
in reactions induced by hadrons and heavy-ions.

%%%%%%%%%%%%%%%%%%%%%%%%%%%%%%%%%%%%%%%%%%%%%%%%%%%%%%%%%%%%%%%%%%%
\section{GiBUU calculations}
\label{sec2}
%%%%%%%%%%%%%%%%%%%%%%%%%%%%%%%%%%%%%%%%%%%%%%%%%%%%%%%%%%%%%%%%%%%

%%%%%%%%%%%%%%%%%%%%%%%%%%%%%%%%%%%%%%%%%%%%%%%%%%%%%%%%%%%%%%%%%%%
\begin{figure}[t]
%\unitlength1cm
%\begin{picture}(10.,8.0)
%\put(1.,0.){\makebox{\psfig{file=Fig1.eps,width=12.0cm}}}
%\end{picture}
\includegraphics[width=.7\textwidth]{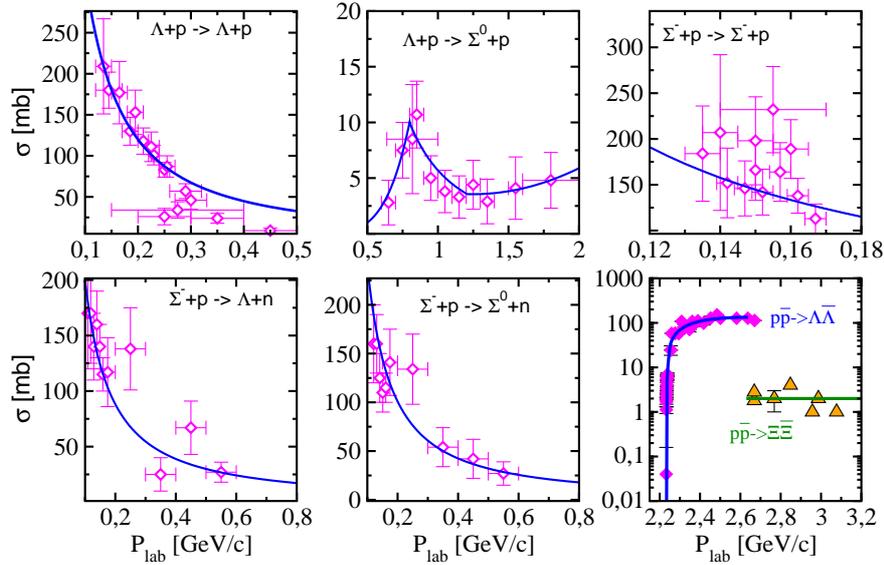}
\caption{Cross sections as function of beam momentum for 
various channels (as indicated) relevant for the 
production of hypernuclei. Theoretical parametrizations (solid curves) 
are compared with data (symbols, taken from~\cite{data}).
}
\label{Fig1}
\end{figure}
%%%%%%%%%%%%%%%%%%%%%%%%%%%%%%%%%%%%%%%%%%%%%%%%%%%%%%%%%%%%%%%%%%%

The fast non-equilibrium stage of reactions is modelled 
by the Giessen-Boltzmann-Uehling-Uhlenbeck (GiBUU) transport
model~\cite{ref3}. The GiBUU model is based on the following 
Boltzmann-Uehling-Uhlenbeck equation
%%%%%%%%%%%%%%%
\begin{equation}
\left[
k^{*\mu} \partial_{\mu}^{x} + \left( k^{*}_{\nu} F^{\mu\nu}
+ m^{*} \partial_{x}^{\mu} m^{*}  \right)
\partial_{\mu}^{k^{*}}
\right] f(x,k^{*}) = {\cal I}_{coll}
\quad .
\label{rbuu}
\end{equation}
%%%%%%%%%%%%%%%
Eq.~(\ref{rbuu}) describes the dynamical evolution of the one-body phase-space 
distribution function $f(x,k^{*})$ for the hadrons under the influence of a 
hadronic mean-field (l.h.s. of Eq.~(\ref{rbuu})) and binary collisions 
(r.h.s. of Eq.~(\ref{rbuu})). Our previous Dirac-Brueckner studies of 
$NN$ and $YN$ in-medium 
interactions~\cite{DBHF} are showing that the real parts of the self-energies 
are well described in Hartree-approximation, at least at low energies. Hence, guided 
by those results, we obtain a numerically feasible approach by extending the 
mean-field approximation to dispersive processes with imaginary parts derived by 
a Monte Carlo procedure (see below). The relativistic mean-field enters into GiBUU 
through the kinetic $4$-momenta $k^{*\mu}=k^{\mu}-\Sigma^{\mu}$ and the 
effective (Dirac) masses $m^{*}=M-\Sigma_{s}$, with the vector and scalar 
self-energy components given by 
$\Sigma^{\mu} = g_{\omega}\omega^{\mu} + \tau_{3}g_{\rho}\rho_{3}^{\mu}$, 
$\Sigma_{s} = g_{\sigma}\sigma$, 
%%%%%%%%%%%%%%%%
%\begin{equation}
%\Sigma^{\mu} = g_{\omega}\omega^{\mu} + \tau_{3}g_{\rho}\rho_{3}^{\mu}
%~~,~~
%\Sigma_{s} = g_{\sigma}\sigma
%\quad ,
%\label{SelfEnergies} 
%\end{equation} 
%%%%%%%%%%%%%%%% 
with the isoscalar, scalar $\sigma$, the isoscalar, vector $\omega^{\mu}$ 
meson fields and the third isospin-component of the isovector, vector $\rho_{3}^{\mu}$
meson field. The meson-nucleon coupling constants $g_{i}$ 
($i=\sigma,\omega,\rho$) and also the additional parameters of the non-linear 
self-interactions of the $\sigma$ meson (not shown here) are taken from the 
widely used NL parametrizations, see~\cite{ref2} for details. 
The antinucleon-meson coupling constants are rescaled by 
a phenomenological constant such, to reproduce data in 
high-energy antiproton-induced reactions~\cite{ref2}.

Furthermore, the collision term is treated in a parallel ensemble 
algorithm incorporating the standard parametrizations for the cross sections 
of various binary processes~\cite{ref2,ref3}. 
For the present calculations we follow the usual scheme of using empirical 
cross section data. The approach discussed in the next section will allow us to 
overcome that limitation in the near future.

GiBUU results on fragment formation 
and on multi-strange $\Lambda$-hypernuclei at PANDA have 
been published in Refs.~\cite{ref2}. Except of a proper description of 
fragmentation~\cite{ref2}, 
the theoretical description of the various elementary channels for  
strangeness production is also important. Fig.~\ref{Fig1} shows some 
examples for secondary and primary hyperon production channels, 
where one realizes the need of more precise measurements of such 
elementary input. Note that not for all the relevant channels data 
exist. In fact, the channel $\Xi N\rightarrow \Lambda\Lambda$, 
which is crucial for the formation of double-$\Lambda$ hypernuclei 
at PANDA, completely relies on theoretical estimates~\cite{ref2}. 

\begin{figure}[t]
%\unitlength1cm
%\begin{picture}(10.,7.0)
%\put(0.75,0.){\makebox{\psfig{file=Fig2.eps,width=12.0cm}}}
%\end{picture}
\includegraphics[width=.6\textwidth]{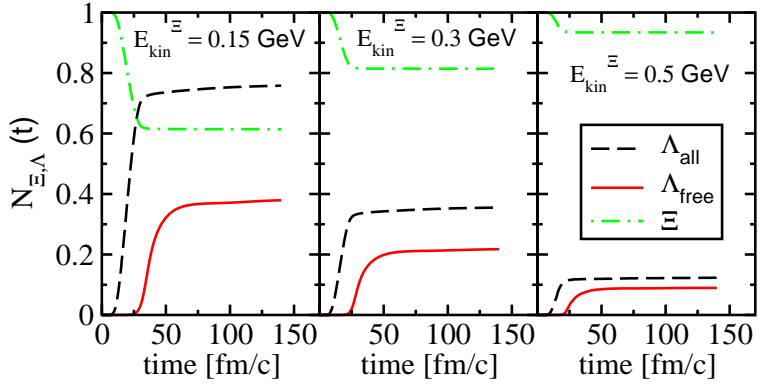}
\caption{GiBUU results for the time dependence of all and free $\Lambda$'s 
and of $\Xi$'s for central $\Xi$-induced reactions on ${}^{64}$Cu-target 
at beam energies as indicated.
}
\label{Fig2}
\end{figure}

Fig.~\ref{Fig2} shows GiBUU results for the time evolution of strangeness 
with $S=-1$ and $S=-2$, for $\Xi$-induced reactions at three $\Xi$-energies. 
The production of bound $\Lambda$ particles strongly decreases with 
$\Xi$-energy. This is due to the enhanced repulsive character of the
mean-field at large positive energies, as typical for baryon 
mean-field self-energies, but also due to the strong decrease of the 
elementary $\Xi N\rightarrow \Lambda\Lambda$ cross section with rising 
energy. These results are compatible with our previous studies~\cite{ref2}.

%%%%%%%%%%%%%%%%%%%%%%%%%%%%%%%%%%%%%%%%%%%%%%%%%%%%%%%%%%%%%%%%%%%
\section{In-medium hyperon-nucleon calculations}
\label{sec3}
%%%%%%%%%%%%%%%%%%%%%%%%%%%%%%%%%%%%%%%%%%%%%%%%%%%%%%%%%%%%%%%%%%%
Reactions induced by (anti)baryons and heavy-ions take place not in 
free space, but inside a dense hadronic environment. Thus, the in-medium 
dependence of the same elementary channels is indispensable for the 
production of hypernuclei, apart from the mean-field effects. First studies 
on this direction have been already started within the Bethe-Salpeter
formalism for $YN$ in-medium scattering. This is done by solving the 
Lippmann-Schwinger (LS) equation for the R-matrix, given in short-hand 
notation by 
%%%%%%%%%%%%%%%
\begin{equation}
R = V + {\cal P}\,\int \, V\, G \, Q \, R
\quad ,
\end{equation}
%%%%%%%%%%%%%%%
where $V$ is the underlying $YN$-potential in free space, 
$G$ the propagator for intermediate states and $Q$ the Pauli operator for 
intermediate states. The integration is performed over intermediate states 
and ${\cal P}$ denotes the principal value. 
Since we account for the coupling of the various $YN$ channels of same 
strangeness $S$ and total charge, the LS equation, in fact, has a matrix 
structure. The free space $V_{YN}$ potential is obtained by rescaling 
the $NN$ coupling constants of the Bonn potential by fits to the scarce 
data quoted above and results by the Juelich group~\cite{ref4,julich}. 
The parameters are determined such that one first starts at low 
energies where $\Lambda\Sigma$ coupling does not occur, and obtains 
the rescaling parameter $b_{\Lambda N}$ for the 
simple case of $\Lambda N \rightarrow \Lambda N$. 
The remaining parameters are determined at higher energies above the 
kinematical thresholds of the channels 
$\Sigma N \rightarrow \Sigma N$ and $\Lambda N \rightarrow \Sigma N$. This 
results in a reduction of the NN coupling parameters by factors 
$b_{\Lambda N, \Sigma N}\simeq 0.8-0.86$ for $\Lambda N \rightarrow \Lambda N$ and 
$\Sigma N \rightarrow \Sigma N$, and $b_{\Lambda\Sigma}=0.02$ for 
$\Lambda N \rightarrow \Sigma N$. Finally, in-medium effects are incorporated 
by solving the LS equation with the inclusion of Pauli projector $Q$. 
Pauli-blocking is in fact the leading order source of medium effects. Since 
the self-energies are contributing only from second order on, they have been 
left out at this moment.

%%%%%%%%%%%%%%%%%%%%%%%%%%%%%%%%%%%%%%%%%%%%%%%%%%%%%%%%%%%%%%%%%%%
%%%%%%%%%%%%%%%%%%%%%%%%%%%%%%%%%%%%%%%%%%%%%%%%%%%%%%%%%%%%%%%%%%%

\begin{figure}[t]
%\unitlength1cm
%\begin{picture}(10.,6.0)
%\put(-0.25,0.){\makebox{\psfig{file=Fig3.eps,width=14.5cm}}}
%\end{picture}
\includegraphics[width=.75\textwidth]{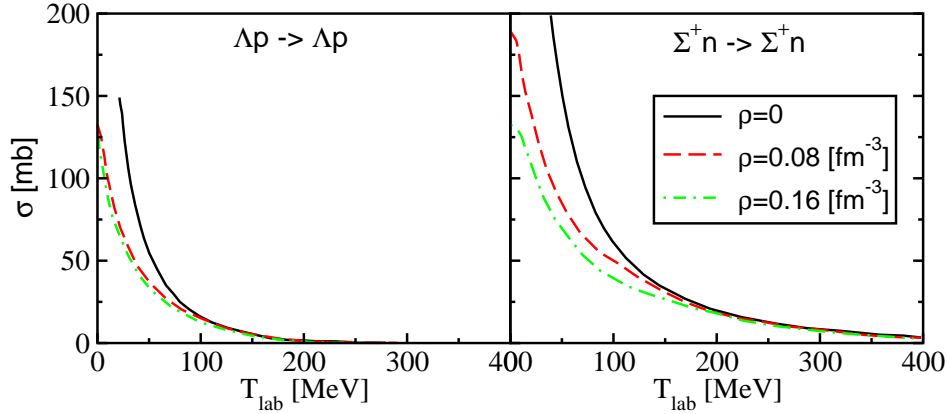}
\caption{Energy dependence of elastic in-medium cross sections 
for $\Lambda N$ and $\Sigma N$ scattering  (as indicated) at densities 
of $\rho=0,~0.08,~0.16~[fm^{-3}]$ (solid, dashed and dot-dashed, 
respectively)~\cite{ref4}.
}
\label{Fig3}
\end{figure}

Fig.~\ref{Fig3} shows the energy dependence of $YN$ elastic in-medium 
cross sections at different densities. 
Inside nuclear matter, the $YN$ interaction strength is further reduced, 
mainly by Pauli-blocking of the intermediate nucleon states. 
As seen from Fig.~\ref{Fig3}, the reduction increases with density, 
leading at saturation density to a suppression of the cross sections by 
about $50\%$. Our results indicate that one must expect important 
in-medium effects of $YN$ secondary scattering in reactions induced by 
heavy-ions (HypHI experiment~\cite{ref5}) and by antiprotons 
(PANDA experiment~\cite{ref1a}) and, therefore, on the production of 
hypernuclei. 
For a proper treatment in transport-theoretical calculations these in-medium 
dependences have to be implemented in transport codes, 
by parametrizing the density and energy dependence of the in-medium 
cross sections. This is feasible and presently in progress.
%%%%%%%%%%%%%%%%%%%%%%%%%%%%%%%%%%%%%%%%%%%%%%%%%%%%%%%%%%%%%%%%%%%
%%%%%%%%%%%%%%%%%%%%%%%%%%%%%%%%%%%%%%%%%%%%%%%%%%%%%%%%%%%%%%%%%%%

\begin{figure}[t]
%\unitlength1cm
%\begin{picture}(10.,6.0)
%\put(-0.25,0.){\makebox{\psfig{file=Fig3.eps,width=14.5cm}}}
%\end{picture}
\includegraphics[width=.5\textwidth]{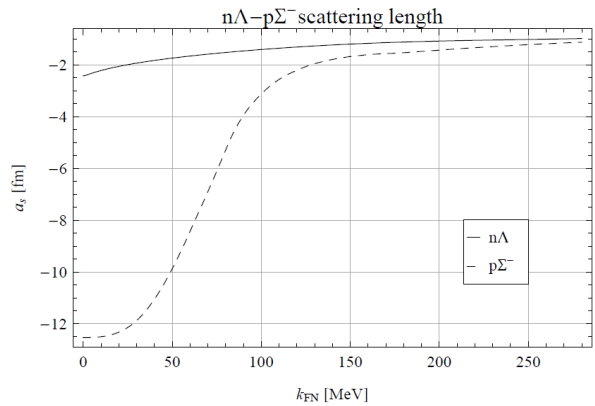}
\caption{Scattering length $a_{s}$ for the channels as indicated as 
function of Fermi-momentum of the nucleonic medium.
}
\label{Fig4}
\end{figure}

A suitable measure for the interaction strength at threshold 
is obtained from the 
low-energy effective range expansion of the $l=0$ phase shift 
%%%%%%%%%%%%%%%
\begin{equation}
\frac{k}{\tan{\delta}} = -\frac{1}{a_{s}} + \frac{1}{2}k^{2}r_{s}
\end{equation}
%%%%%%%%%%%%%%%
leading to the effective range $r_{s}$ and to the scattering length 
$a_{s}$, which is proportional to the volume integral of the underlying 
$YN$ potential. In Fig.~\ref{Fig4}, $a_{s}$ is 
displayed for the $n\Lambda$ and $p\Sigma^{-}$ systems as function 
of $k=k_{F}$, i.e., the on-shell momentum is taken at the 
Fermi-momentum of the nucleonic background medium. Close to 
saturation a strong reduction, especially in the $p\Sigma^{-}$-channel, 
is observed, approaching the strength of the $n\Lambda$-channel.

%%%%%%%%%%%%%%%%%%%%%%%%%%%%%%%%%%%%%%%%%%%%%%%%%%%%%%%%%%%%%%%%%%%
\section{Final remarks}
\label{sec4}
%%%%%%%%%%%%%%%%%%%%%%%%%%%%%%%%%%%%%%%%%%%%%%%%%%%%%%%%%%%%%%%%%%%
In summary, the production of hypernuclei in reactions induced by heavy-ions 
and (anti)protons opens the opportunity to explore the less known hyperon-nucleon 
and hyperon-hyperon interactions at densities close to saturation and beyond. 
First transport-theoretical studies within the GiBUU transport model have been 
already performed. Here the in-medium effects of various channels relevant for 
the production of hypernuclei have been studied and have been found to be important, 
in particular, at low energies. Since the hypernuclear formation takes place inside a rather 
highly excited hadronic environment, possible in-medium effects, which emerge at 
finite baryon density, have to be taken into account in transport simulations of 
reactions.

%%%%%%%%%%%%%%%%%%%%%%%%%%%%%%%%%%%%%%%%%%%%%%%%%%%%%%%%%%%%%%%%%%%
%%%%%%%%%%%%%%%%%%%%%%%%%%%%%%%%%%%%%%%%%%%%%%%%%%%%%%%%%%%%%%%%%%%

\section*{Acknowledgments}
%{\it Acknowledgments.} 
This work is supported 
by BMBF contract GILENS 06, DFG contract Le439/9-1 and HIC for FAIR.

%\section*{References}
%%%%%%%%%%%%%%%%%%%%%%%%%%%%%%%%%%%%%%%%%%%%%%%%%%%%%%%%%%%%%%%%%%%%%%%%%
%                                                                       %
%   BEGIN OF BIBLIOGRAPHY                                               %
%                                                                       %
%%%%%%%%%%%%%%%%%%%%%%%%%%%%%%%%%%%%%%%%%%%%%%%%%%%%%%%%%%%%%%%%%%%%%%%%%
%\begin{thebibliography}{99}

\end{document}